\documentclass[sigconf]{acmart}

\usepackage{color}

\usepackage{multirow}
\usepackage{mathrsfs}
\usepackage{dsfont}
\usepackage{amsmath}
\usepackage{algpseudocode}
\usepackage{subcaption}
\usepackage{booktabs}
\usepackage{bm}
\usepackage{algorithm}
\usepackage{hyperref}

\AtBeginDocument{%
  \providecommand\BibTeX{{%
    \normalfont B\kern-0.5em{\scshape i\kern-0.25em b}\kern-0.8em\TeX}}}

\copyrightyear{2022}
\acmYear{2022}
\setcopyright{acmlicensed}\acmConference[MM '22]{Proceedings of the 30th
ACM International Conference on Multimedia}{October 10--14, 2022}{Lisboa,
Portugal}
\acmBooktitle{Proceedings of the 30th ACM International Conference on
Multimedia (MM '22), October 10--14, 2022, Lisboa, Portugal}
\acmPrice{15.00}
\acmDOI{10.1145/3503161.3548084}
\acmISBN{978-1-4503-9203-7/22/10}

\begin{document}

\title{Structure-Enhanced Pop Music Generation via Harmony-Aware Learning}



\author{Xueyao Zhang}
\authornote{This work was accomplished when Xueyao Zhang and Li Wang worked as interns at Pattern Recognition Center, WeChat AI, Tencent Inc.}
\affiliation{%
  \institution{The Chinese University of Hong Kong, Shenzhen}
  \city{}\country{}
 }
\email{zhangxueyao1998@gmail.com}

\author{Jinchao Zhang}
\authornote{Corresponding author.}
\affiliation{%
  \institution{Pattern Recognition Center, WeChat AI, Tencent Inc}
\city{}\country{}
 }
\email{dayerzhang@tencent.com}

\author{Yao Qiu}
\affiliation{%
  \institution{Pattern Recognition Center, WeChat AI, Tencent Inc}
\city{}\country{}
 }
\email{yasinqiu@tencent.com}

\author{Li Wang}
\authornotemark[1]
\affiliation{%
  \institution{Communication University of China}
\city{}\country{}
 }
\email{wwli@cuc.edu.cn}

\author{Jie Zhou}
\affiliation{%
  \institution{Pattern Recognition Center, WeChat AI, Tencent Inc}
\city{}\country{}
 }
\email{withtomzhou@tencent.com}



\begin{abstract}
  Pop music generation has always been an attractive topic for both musicians and scientists for a long time. However, automatically composing pop music with a satisfactory \textit{structure} is still a challenging issue. In this paper, we propose to leverage \textit{harmony}-aware learning for structure-enhanced pop music generation. On the one hand, one of the participants of harmony, \textit{chord}, represents the harmonic set of multiple notes, which is integrated closely with the spatial structure of music, the \textit{texture}. On the other hand, the other participant of harmony, \textit{chord progression}, usually accompanies the development of the music, which promotes the temporal structure of music, the \textit{form}. Moreover, when chords evolve into chord progression, the texture and form can be bridged by the harmony naturally, which contributes to the joint learning of the two structures. Furthermore, we propose the \underline{H}armony-\underline{A}ware Hierarchical Music \underline{T}ransformer (HAT), which can exploit the structure adaptively from the music, and make the musical tokens interact hierarchically to enhance the structure in multi-level musical elements. Experimental results reveal that compared to the existing methods, HAT owns a much better understanding of the structure and it can also improve the quality of generated music, especially in the \textit{form} and \textit{texture}.\footnote{The code and generated pieces are available at \href{https://github.com/RMSnow/HAT}{https://github.com/RMSnow/HAT}.}
\end{abstract}




\maketitle

\section{Introduction}\label{sec:intro}

\begin{figure*}[ht]
     \centering
     \begin{subfigure}[b]{0.8\textwidth}
         \centering
         \includegraphics[width=\textwidth]{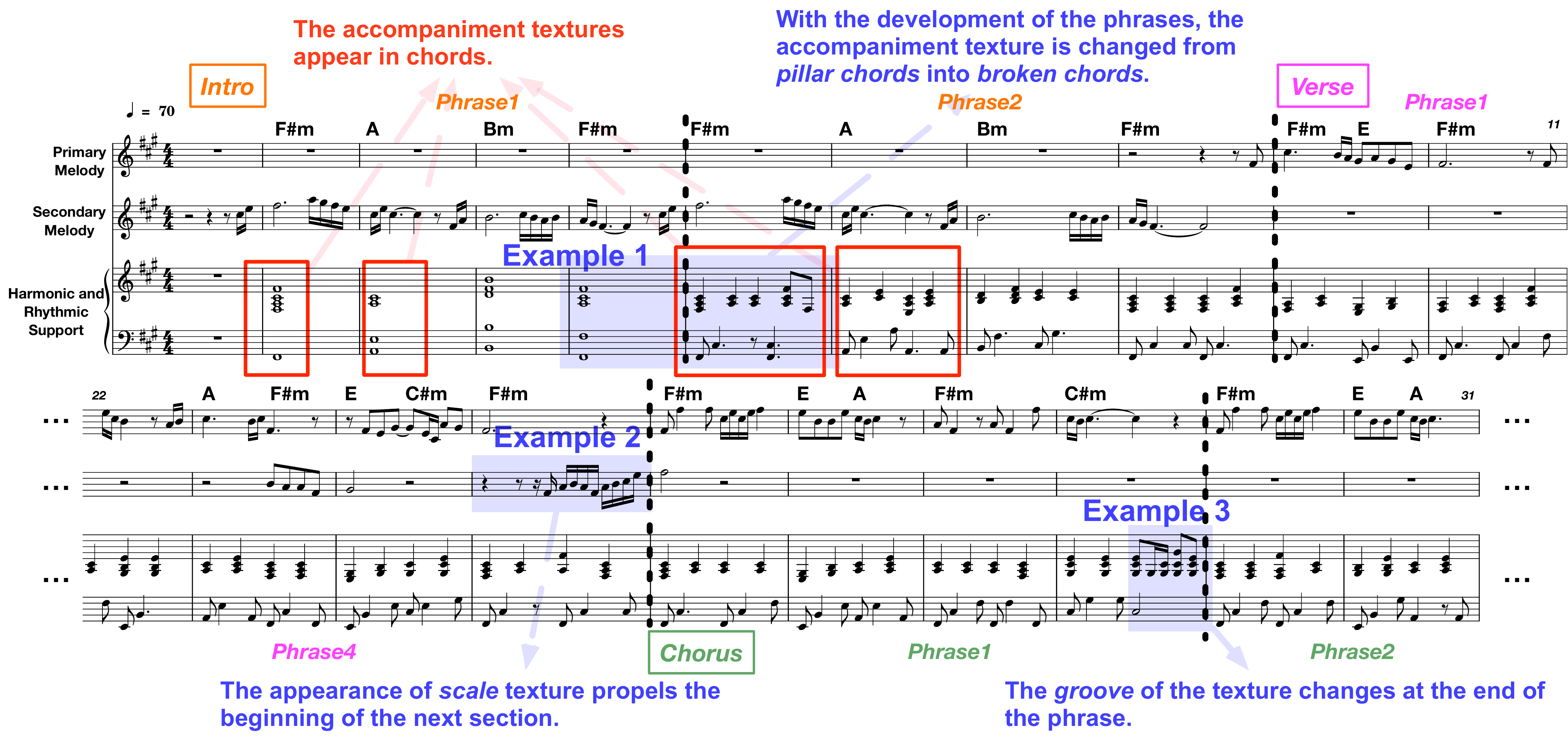}
         \caption{Part of the score.}
         \label{fig:intro-case-score}
     \end{subfigure}
     \hfill
     \begin{subfigure}[b]{0.7\textwidth}
         \centering
         \includegraphics[width=\textwidth]{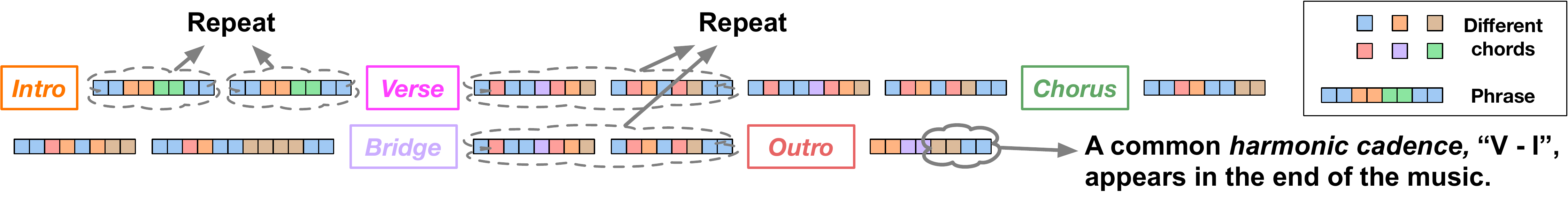}
         \caption{The \textit{chord progressions} of each \textit{phrases} within the \textit{sections}.}
         \label{fig:intro-case-cp}
     \end{subfigure}
\caption{A motivating example: the Chinese pop song, \textit{Yi Jian Mei}. It reveals that: (1) The \textit{chord} is integrated closely with the \textit{texture} structure (see the red blocks of Figure~\ref{fig:intro-case-score}); (2) The \textit{chord progression} accompanies with the formation of the \textit{form} structure (Figure~\ref{fig:intro-case-cp}); (3) There is a highly mutual dependency between the \textit{texture} and the \textit{form} (see the blue blocks of Figure~\ref{fig:intro-case-score}).}
\label{fig:intro-case}
\end{figure*}

Composing music by computational means (also known as \textit{algorithmic composition} or \textit{automatic music generation}) is a longstanding human desire to explore the frontiers of computational creativity~\cite{computational-creativity-2020}. 
As an essential attribute to music, \textit{structure} is inherently tied to human perception and cognition~\cite{structure-psychological-1990,structure-perception-2006}, but is hard for the computer to recognize and identify~\cite{FMP-book-2015,structure-human-perceive-2016}. Moreover, automatically composing a music with balanced, coherent, and integrated structure is an attractive, such as the early researches tens of years ago~\cite{EMI-1987,Todd-1989}, but still challenging issue~\cite{survey-taxonomy-2017,briot-book-2020}. 


In musicology, usually we can analyze the \textit{structure} from two aspects, \textit{form} and \textit{texture}. Horizontally, \textit{form} reveals the temporal relationship and dependency among the music, such as the repetition of motives, the transition between phrases, and the development between sections~\cite{wiki-form}. Vertically, \textit{texture} represents the spatial relationship and the organized way between the multiple parts or instruments of music~\cite{wiki-texture}. For example, the typical \textit{texture} of pop music is that the melody stands out prominently and the others form a background of harmonic accompaniment, i.e. \textit{homophony}~\cite{wiki-homophony}. 

Overall, the characteristics of the musical structure lie in the three aspects: (1) Firstly, the structure depends largely on the musical context and is hard to be described clearly and defined accurately~\cite{FMP-book-2015}. (2) Secondly, the structure exists in various musical elements and appears the hierarchy, ranging from the low-level \textit{motif} to the high-level \textit{phrase} and \textit{section}~\cite{dai-structure-analysis-2020}. (3) Thirdly, the \textit{form} and the \textit{texture} are connected closely and support to each other. Specifically, it is common in pop music that the \textit{texture} is consistent within a specific phrase or section, while is changed with the development of the \textit{form}. For example, in Figure~\ref{fig:intro-case-score}, the accompaniment texture is \textit{pillar chords} in the first phrase of the intro, while is changed into \textit{broken chords} in the second phrase (as the blue solid block ``Example 1" shows). The mutual dependency can be also observed in ``Example 2" and ``Example 3". 

Therefore, a model aiming to produce well-structured music should meet the three corresponding requirements:
 
 \begin{itemize}
 	\item \textbf{R1}: It should mine the contextual pattern of structure from the music data adaptively.
 	\item \textbf{R2}: It should exploit the appropriate musical elements to represent structure units.
 	\item \textbf{R3}: It should capture the highly mutual dependency between \textit{form} and \textit{texture}.
 \end{itemize}
 \vspace{-2pt}
 
Based on the aforementioned requirements, we propose to leverage \textit{harmony}-aware learning for structure-enhanced pop music generation. In musicology, the study of \textit{harmony} involves \textit{chords} and their construction, and \textit{chord progressions} and the principles of connection that govern them~\cite{harmony-dictionary}. The reasons to learn harmony are that for \textbf{R1}, harmony represents the consonance of the musical context, so we can mine the musical contextual information by learning it. For \textbf{R2}, not only harmony itself is an important structure element, but also it combines the many musical elements organically, from the low-level notes to the high-level phrase and sections. Specifically, one of the participants of harmony, \textit{chord}, represents the harmonic set of multiple notes, which is integrated closely with the \textit{texture}. As the red hollow blocks in Figure~\ref{fig:intro-case-score} show, the accompaniment textures always appear in chords. Besides, the other participant, \textit{chord progression}, usually promotes the development of the music, contributing to the formation of the \textit{form} (such as the \textit{harmonic cadences}~\cite{cadence-1999} in Figure~\ref{fig:intro-case-cp}). 
Moreover, for \textbf{R3}, we can model the evolution from \textit{chord} to \textit{chord progression} to bridge between \textit{texture} and \textit{form}, which reveals the feasibility to the joint learning of the two structures.

In the paper, our contributions are summarized as follows:

\begin{itemize}
	\item We propose to learn the musical prior knowledge, \textit{harmony}, for structure-enhanced pop music generation. To the best of our knowledge, our work is the first to learn \textit{form} and \textit{texture} jointly bridged by \textit{harmony} in algorithmic composition.
	\item We design an end-to-end model, \underline{H}armony-\underline{A}ware Hierarchical Music \underline{T}ransformer (HAT), to produce well-structured pop music. It can model the musical structure by rendering musical elements interact at the hierarchical levels.
	\item We develop two objective metrics for evaluating the \textit{structure} of music from the perspective of the \textit{harmony}.
	\item Experimental results verify the effectiveness of HAT on both music understanding and generation, especially in the \textit{form} and \textit{texture}.
\end{itemize}
 \vspace{-6pt}



 
\vspace{-3pt}
\section{Related Work}

\subsection{Structure-Enhanced Music Generation}
Improving the quality of the musical \textit{structure} is always the focus in music generation. Because of the difficulty of modeling structure, some researchers deal with the task in pipelines with multiple models~\cite{multiple-models-1998,PopMNet-2020,music-framework-ismir21}, like designing a single model to modulate the structure and the others to generate the music based on it. For the end-to-end methods, quite a few researchers consider the structure at only high-level like \textit{section}, and still use templated- or rule-based methods. For example, Zhou et al. use a predefined \textit{section} sequence, such as "AABA", to generate a structured music~\cite{BandNet-2019}. And others utilize the \textit{chord progressions} as the constraint rules, making them as explicit input to generate structure-enhanced music~\cite{wangziyu-cp-framework-2018,XiaoIceBand-2018,explicit-structure-2019}.

As for the end-to-end generators mining the structure adaptively, in StructureNet~\cite{StructureNet-2018}, the authors leverage the RNN to exploit the structural repeat in the music. However, the definition of structure in~\cite{StructureNet-2018} is simple (just two types of repeat), and the StructureNet focuses on producing the monophony music, whose texture is only a single melodic line. In MusicVAE~\cite{MusicVAE-2018} and TransformerVAE~\cite{TransformerVAE-2020}, the authors aim to mine the structure dependency between the \textit{bars}. However, \textit{bar} as a structure unit is not flexible, compared to other elements such as \textit{motif}, \textit{phrase} or \textit{section}. Besides, the generators of the two are not capable of producing the song-length music. In Music Transformer~\cite{MusicTransformer-2019}, the researchers introduce the relative attention to model the position relationships between the \textit{notes}, hoping to learn the long-term structure at the \textit{note} level. However, it lacks the explicit modeling for more abstract structure including the transition and the development of \textit{phrases} and \textit{sections}. Recently, the designs of most advanced generation models are Transformer-architecture~\cite{PopMusicTransformer-REMI-2020,CP-Transformer-2021,taiwan-arxiv-2021}. Most of them leverage the self ability of the Transformer to learn the long-term dependency, but there is almost none explicit modeling for the structure-enhanced generation.

\subsection{Harmony Learning} 
In the field of music representation learning, there are researches aiming to learn the representation of the harmony~\cite{PianoTree-2020,wangziyu-pianotree2-2020}. Nevertheless, these works focus more on representation learning than producing well-structured music pieces. Moreover, the generators of these works are designed to learn just several \textit{bars} of music, but hard to handle the \textit{song-length} music. 

In a word, for structure-enhanced pop music generation, our proposed HAT is the first end-to-end generator that learns \textit{form} and \textit{texture} jointly bridged by \textit{harmony} to the best of our knowledge.


\section{Methodology}\label{sec:method}

\begin{figure*}[ht]
     \centering
     \begin{subfigure}[b]{0.31\textwidth}
         \centering
         \includegraphics[width=\textwidth]{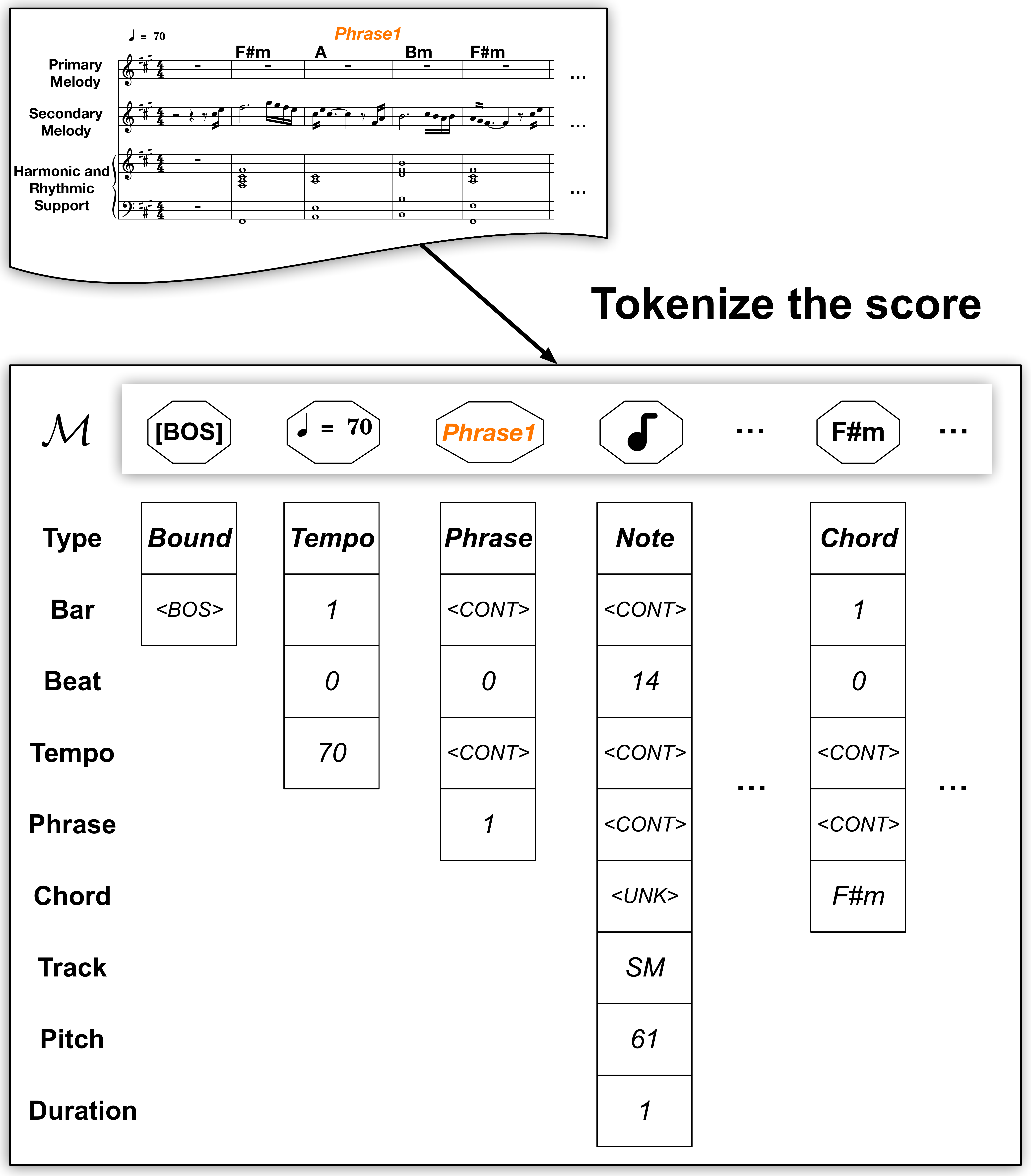}
         \caption{Music Tokenization. }
         \label{fig:HAT-tokenize}
     \end{subfigure}
     \hfill
     \begin{subfigure}[b]{0.68\textwidth}
         \centering
         \includegraphics[width=\textwidth]{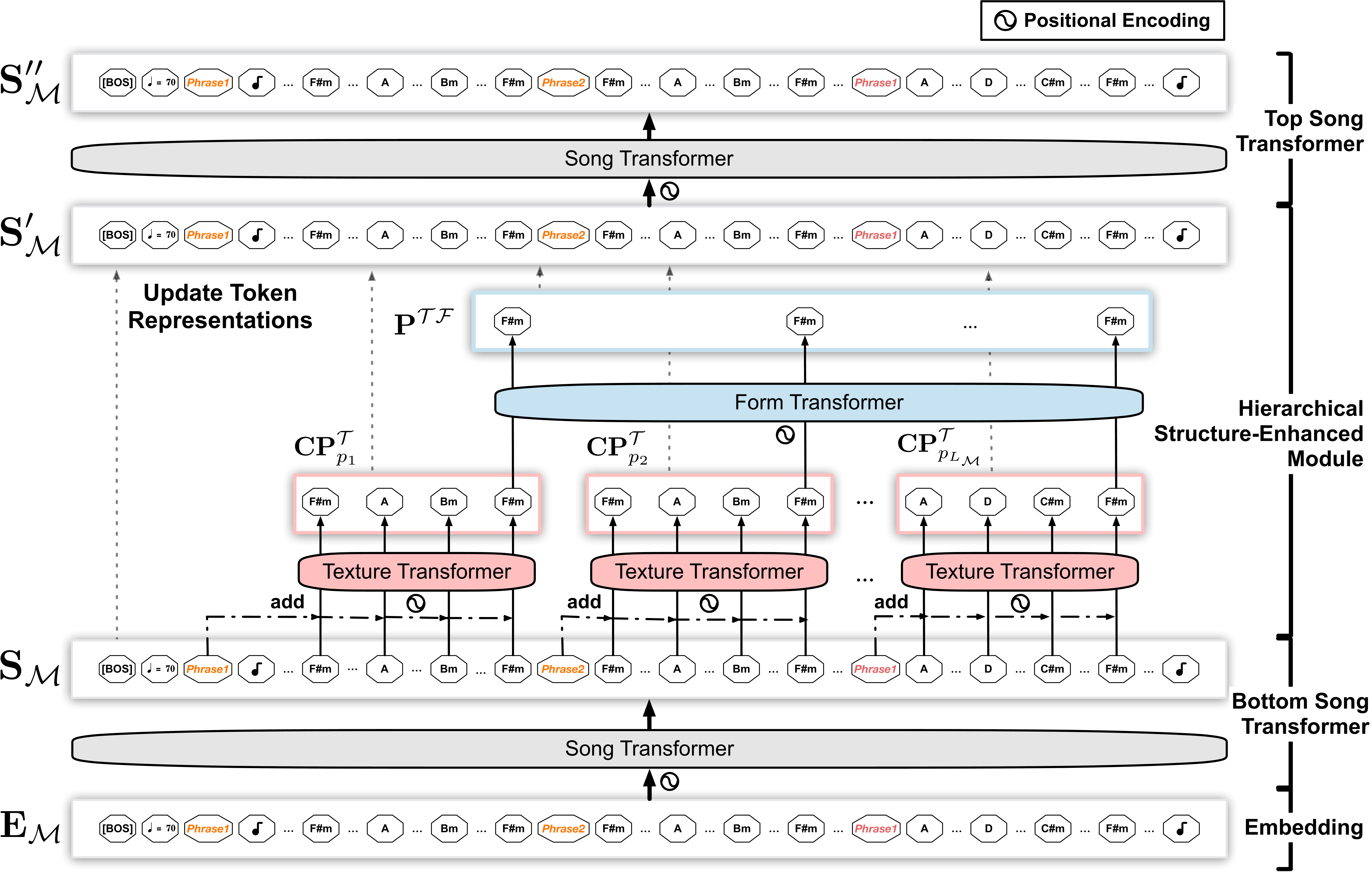}
         \caption{Musical tokens interact at the hierarchical levels.}
         \label{fig:HAT-model}
     \end{subfigure}
     \caption{The architecture of Harmony-Aware Hierarchical Music Transformer (HAT). Firstly, we tokenize the symbolic music, where every token is mapped to nine events (Figure~\ref{fig:HAT-tokenize}). Then, we use the bottom Song Transformer, the Hierarchical Structure-Enhanced Module, and the top Song Transformer to obtain the structure-enhanced tokens (Figure~\ref{fig:HAT-model}). }
     \label{fig:HAT}
\end{figure*}

\subsection{Overview}\label{sec:model-overview}


%

In the paper, we propose the Harmony-Aware Hierarchical Music Transformer (HAT) for structure-enhanced pop music generation (Figure~\ref{fig:HAT}). Firstly, we adopt the event-based tokenization to represent the symbolic music data as input (Figure~\ref{fig:HAT-tokenize}). Subsequently, we utilize the three Transformer-based~\cite{Transformer-2017} blocks (Song Transformer, Texture Transformer, and Form Transformer) to make the tokens interact at the hierarchical levels (Figure~\ref{fig:HAT-model}). 

Aiming to capture the mutual dependency between \textit{form} and \textit{texture}, we design the hierarchical structure-enhanced mechanisms (Hierarchical Structure-Enhanced Module, HSE) in particular (Figure~\ref{fig:HAT-model}). As is described in Section~\ref{sec:intro}, usually the \textit{textures} \textbf{stay within \textit{phrases}}, but \textbf{change on their boundaries}. To model this characteristic, therefore, firstly we group the chord tokens within every phrase, which means the chord progression of the current phrase, and use Texture Transformer to learn the \textit{local texture}. Then we input the texture of every phrase's last chord into Form Transformer to learn the \textit{global form}. 

The workflow of HAT is as follows. During training: 


\begin{enumerate}
	\item The bottom Song Transformer handles all the musical tokens.
	\item The HSE module treat chords and phrases as the special structure indicators and update their representations.
	\item The top Song Transformer make all the tokens interact again, which can broadcast the explored information of structure to the multi-grained elements.
	\item Do the predictions by the structure-enhanced tokens.
\end{enumerate}

During the generation, the HAT can produce pieces from scratch or with the guidance of the specific prompts.

\subsection{Music Tokenization}\label{sec:music-tokenization}

In order to represent the symbolic music, we adopt the event-based tokenization (Figure~\ref{fig:HAT-tokenize}). It can serialize the multi-type musical information into a sequence of one-hot encoded events~\cite{event-based-tokenize-2020}, which is applied broadly for music generation~\cite{MusicTransformer-2019,PopMusicTransformer-REMI-2020,CP-Transformer-2021}. 

\begin{table}[h]
\centering
\caption{Nine events in music tokenization.}
\label{tab:tokenize}
\resizebox{\columnwidth}{!}{%
\begin{tabular}{ccl}
\toprule
\textbf{Level} & \textbf{Event} & \textbf{Description} \\ \midrule
Token type & Type & The type of the token \\ \midrule
\multirow{3}{*}{Metrical} & Bar & The bar position of the token \\
 & Beat & The beat position in a bar of the token \\
 & Tempo & The tempo of the token \\ \midrule
\multirow{2}{*}{Structure} & Phrase & The phrase that the token belongs with \\
 & Chord & The chord that the token belongs with \\ \midrule 
\multirow{3}{*}{Note} & Track & The track (or the instrument) of the token \\
 & Pitch & The pitch of the token \\
 & Duraion & The duration time of the token \\ 
 \bottomrule 
\end{tabular}%
}
\end{table}

We employ an nine-event set $\mathcal{C}$ (Table~\ref{tab:tokenize}).\footnote{In this paper, we focus on producing the \textit{symbolic} music, so we leave the \textit{performance} musical attributes like "velocity" for the future research, i.e. \textit{performance} generation.} Specially, it contains two structure-level events, Phrase and Chord, which will be the structure-enhanced indicators in HAT.

Given the music $\mathcal{M} = [t_1, t_2, ..., t_L]$, where $t_i$ is the $i^{th}$ token and $L$ is the length of the music. For the token $t_i$, we embed its each event values and concatenate them to obtain $\mathbf{E}_{t_i}$:

\begin{equation}
	\mathbf{E}_{t_i} = {\rm \textbf{Concat}}([{\rm \textbf{Emb}}_c(t_i^c), {\rm for~every~} c \in \mathcal{C}]),
\label{eq:embedding}
\end{equation}
where {\rm \textbf{Concat}} means the concatenation, $t_i^c$ is the value of $t_i$ on the category $c$, and $\textbf{Emb}_c(\cdot)$ is the embedding layer for the category $c$. 

\subsection{Bottom Song Transformer}
To make all the tokens aware of the global musical contexts, we make them interact at the song-level transformer block, the bottom Song Transformer (Figure~\ref{fig:HAT-model}). 

The architecture of Song Transformer is the same as the original Transformer~\cite{Transformer-2017}, except that it is implemented to be autoregressive. Besides, we use the triangular mask strategy of the Transformer Decoder~\cite{Transformer-2017} to guarantee the token at position $i$ can depend only on the tokens at positions less than $i$.

Given the tokenized representation $\mathbf{E}_\mathcal{M} = [\mathbf{E}_{t_0}, \mathbf{E}_{t_1}, ..., \mathbf{E}_{t_L}]$, we get the updated representation after the bottom Song Transformer, notating $\mathbf{S}_\mathcal{M} = [\mathbf{S}_{t_1}, \mathbf{S}_{t_2}, ..., \mathbf{S}_{t_L}] \in \mathds{R}^{L \times D_{S}}$, where $D_{S}$ is the embedding dimension of Song Transformer:

\begin{equation}
\mathbf{S}_\mathcal{M} = {\rm \textbf{Transformer}}({\rm \textbf{PosEnc}}(\mathbf{E}_\mathcal{M})),
\label{eq:bottom-song}
\end{equation}
where ${\rm \textbf{PosEnc}(\cdot)}$ means the input is added by the sinusoidal position embeddings of Transformer~\cite{Transformer-2017}.

\subsection{Hierarchical Structure-Enhanced Module}
As is mentioned in Section~\ref{sec:model-overview}, the HSE module (Figure~\ref{fig:HAT-model}) is designed for learning the mutual dependency between \textit{form} and \textit{texture}. Specifically, we enhance the attention of \textit{chord} and \textit{phrase} tokens in a hierarchical way in this module.

\subsubsection{Texture Transformer} 
Firstly, we group the \textit{chord} tokens within every \textit{phrase}, which means the \textit{chord progression} of the \textit{phrase}, and use Texture Transformer to learn the local \textit{texture}. The architecture of Texture Transformer is the same as the Song Transformer, except for the hyperparameters such as the number of layers or the number of heads. 

Given the chord progression of the phrase $p_i$, notating $\mathbf{CP}_{p_i} = [\mathbf{S}_{p_i^1}, \mathbf{S}_{p_i^2}, ... , \mathbf{S}_{p_i^{L_{p_i}}}] \in \mathds{R}^{L_{p_i} \times D_{S}}$, where $L_{p_i}$ is the number of chords of the phrase $p_i$, we can obtain the texture-enhanced chords representation after the Texture Transformer, $\mathbf{CP}^{\mathcal{T}}_{p_i} = [\mathbf{T}_{p_i^1}, \mathbf{T}_{p_i^2}, ..., \mathbf{T}_{p_i^{L_{p_i}}}]$:

\begin{equation}
\mathbf{CP}^{\mathcal{T}}_{p_i} = {\rm \textbf{Transformer}}({\rm \textbf{PosEnc}}(\mathbf{S}_{p_i} + \mathbf{CP}_{p_i})).
\label{eq:texture}
\end{equation}

In Equation~\ref{eq:texture}, we add the phrase $p_i$'s representation $\mathbf{S}_{p_i}$ into the chord progression $\mathbf{CP}_{p_i}$ before the Texture Transformer, aiming to fuse the \textit{form}'s dependency when learning \textit{texture}. For the local texture of every phrase, $\mathbf{T}_{p_i}$, we use the last chord of the phrase to represent the local texture, i.e., $\mathbf{T}_{p_i} = \mathbf{T}_{p_i^{L_{p_i}}}$. And we can get the phrases' texture representation, $\mathbf{P}^{\mathcal{T}} = [\mathbf{T}_{p_1}, \mathbf{T}_{p_2}, ..., \mathbf{T}_{p_{L_\mathcal{M}}}] \in \mathds{R}^{L_\mathcal{M} \times D_{S}}$, where $L_\mathcal{M}$ is the phrases num of the music $\mathcal{M}$.

\subsubsection{Form Transformer} 

Next, we aim to learn the global \textit{form} from the local phrase \textit{texture} by Form Transformer. Like Texture Transformer, the architecture of Form Transformer is also homogeneous with the Song Transformer.

Given the phrases' texture representation $\mathbf{P}^{\mathcal{T}}$, we obtain the form-enhanced representation after the Form Transformer, $\mathbf{P}^{\mathcal{TF}} = [\mathbf{TF}_{p_1}, \mathbf{TF}_{p_2}, ..., \mathbf{TF}_{p_{L_\mathcal{M}}}] \in \mathds{R}^{L_\mathcal{M} \times D_{S}}$:

\begin{equation}
	\mathbf{P}^{\mathcal{TF}} = {\rm \textbf{Transformer}}({\rm \textbf{PosEnc}}(\mathbf{P}^{\mathcal{T}})).
\label{eq:form}
\end{equation}

\subsubsection{Update Tokens Representations}

Finally, we update the \textit{chord} and \textit{phrase} tokens, which merges the raw representation with the learned texture-enhanced and form-enhanced representation. Specifically, for the \textit{phrase} tokens (except the first phrase), we add them with the previous both \textit{texture} and \textit{form} context information:

\begin{equation}
		{\mathbf{S}_{p_i}^\prime} = \left\{
			\begin{aligned}
			& \mathbf{S}_{p_i} & i=1, \\
			& \mathbf{TF}_{p_{i-1}} + \mathbf{S}_{p_i} & otherwise.
			\end{aligned}
			\right. \\
\label{eq:update-phrase}
\end{equation}
For the \textit{chord} tokens, we enhance them with their \textit{form} context information. Besides, we also add them with the previous \textit{texture} of their \textit{chord progression} context information (except the first chord of every phrase). Namely, for every $i \in [1, 2, ..., L_\mathcal{M}]$, 

\begin{equation}
		{\mathbf{S}_{p_i^j}^\prime} = \left\{
			\begin{aligned}
			& {\mathbf{S}_{p_i}^\prime} + \mathbf{S}_{p_i^j} & j=1, \\
			& {\mathbf{S}_{p_i}^\prime} + \mathbf{T}_{p_i^{j-1}} + \mathbf{S}_{p_i^j} & otherwise.
			\end{aligned}
			\right. \\
\label{eq:update-chord}
\end{equation}
We leave the other tokens representations unchanged, and obtain the music tokenize representation ${\mathbf{S}_\mathcal{M}^\prime} = [{\mathbf{S}_{t_1}^\prime}, {\mathbf{S}_{t_2}^\prime}, ..., {\mathbf{S}_{t_L}^\prime}] \in \mathds{R}^{L \times D_{S}}$ after the HSE module.


\subsection{Top Song Transformer}
After the HSE module, we have obtained the structure-enhanced chord and phrase tokens. To broadcast the explored structure information into various musical elements of the whole context, we use another Song Transformer block to interact on all the tokens again, getting $\mathbf{S}_\mathcal{M}^{\prime\prime} = [{\mathbf{S}_{t_1}^{\prime\prime}}, {\mathbf{S}_{t_2}^{\prime\prime}}, ..., {\mathbf{S}_{t_L}^{\prime\prime}}] \in \mathds{R}^{L \times D_S}$:


\begin{equation}
	\mathbf{S}_\mathcal{M}^{\prime\prime} = {\rm \textbf{Transformer}}({\rm \textbf{PosEnc}}({\mathbf{S}_\mathcal{M}^\prime})).
\label{eq:top-song}
\end{equation}

\subsection{Training and Generation}\label{sec:train-and-generate}

\subsubsection{Training}
In training, we formulate the music generation as a ``next token prediction" task.
For the music $\mathcal{M} = [t_1, t_2, .., t_L]$, we set the boundary token $t_1 = {\rm \texttt{<BOS>}}$ and $t_{L+1} = {\rm \texttt{<EOS>}}$. Let the input $\mathbf{X} = [\mathbf{x}_1, \mathbf{x}_2, ..., \mathbf{x}_L] = [t_1, t_2, .., t_L]$, and the ground truth $\mathbf{y} = [\mathbf{y}_1, \mathbf{y}_2, ..., \mathbf{y}_L] = [t_2, t_3, .., t_{L+1}]$. we aim to learn a model $G (\mathbf{X}) \rightarrow \hat{\mathbf{y}}$ such that it maximizes the predictive accuracy w.r.t $\mathbf{y}$, where $\hat{\mathbf{y}} = [\hat{\mathbf{y}}_1, \hat{\mathbf{y}}_2, ..., \hat{\mathbf{y}}_L]$.

Given the token representations after the top Song Transformer $\mathbf{S}_\mathcal{M}^{\prime\prime}$, we adopt a two-stage prediction setting to make it easier for the model to fit, following~\cite{CP-Transformer-2021}:


\begin{equation}
\begin{split}
	\hat{\mathbf{y}}_i^{tp} = &{\rm \textbf{Softmax}}({\rm \textbf{MLP}}_{tp} (\mathbf{S}_{t_i}^{\prime\prime})), i=1,2,...,L \\
	\hat{\mathbf{y}}_i^c = &{\rm \textbf{Softmax}}({\rm \textbf{MLP}}_c ({\rm \textbf{Concat}}([\mathbf{S}_{t_i}^{\prime\prime}, {\rm \textbf{Emb}}_{tp} (\mathbf{y}_i^{tp})]))), \\
	& \qquad \qquad \quad i=1,2,...,L;~c \in \mathcal{C}~and~c \neq tp
\end{split}
\label{eq:prediction}
\end{equation}
where $\hat{\mathbf{y}}_i^c$ means the prediction value on the category $c$ of the token $\hat{\mathbf{y}}_i$, $tp$ means the category Type, ${\rm \textbf{Softmax}}(\cdot)$ means the softmax function, ${\rm \textbf{MLP}}_c(\cdot)$ means the Multi-Layer Perceptron on the category $c$, and $\mathbf{y}_i^c$ means the ground truth value on the category $c$ of the token $\mathbf{y}_i$.

During training, we minimize the sum of every cross-entropy loss between the prediction $\hat{\mathbf{y}}$ and the label $\mathbf{y}$ on every category $c$:

\begin{equation}
	\mathcal{L}(\mathbf{y}, \hat{\mathbf{y}}) = \sum_{i=1}^L \sum_{c \in \mathcal{C}} \lambda_c {\rm \textbf{CELoss}}(\mathbf{y}_i^c, \hat{\mathbf{y}}_i^c),
\end{equation}
where $\lambda_c$ is the loss weight on the category $c$, and ${\rm \textbf{CELoss}(\cdot,\cdot)}$ means the cross-entropy loss function.

\subsubsection{Generation}
During the generation, we get the music tokens recurrently. Take the generation from scratch as an example: given the input $\mathbf{x}_1 = {\rm \texttt{<BOS>}}$, the HAT can produce $\hat{\mathbf{y}} = [\hat{\mathbf{y}}_1, \hat{\mathbf{y}}_2, ..., \hat{\mathbf{y}}_{L_G}]$ recurrently, until $\hat{\mathbf{y}}_{L_G} = {\rm \texttt{<EOS>}}$, where $L_G$ is the length of the generated piece. To obtain the final predictive music tokens $[\hat{t_1}, \hat{t}_2, ..., \hat{t}_{L_G}]$, we adopt the stochastic temperature controlled sampling~\cite{sampling-strategy-2020} to increase the diversity and avoid degeneration:

\begin{equation}
\hat{t}_i^c = {\rm \textbf{Sampling}}_c(\hat{\mathbf{y}}_i^c),
\end{equation}
where $\hat{t}_i^c$ is the value of predictive token $\hat{t}_i$ on the category $c$, and ${\rm \textbf{Sampling}}_c(\cdot)$ means the sampling function on the category $c$. Here we employ different sampling policies for different categories, following~\cite{CP-Transformer-2021}.


\section{Experiments}
In the section, we conduct experiments to answer the following evaluation questions:

\textbf{EQ1:} Does HAT have a better perception and understanding of the musical structure compared to the existing models?

\textbf{EQ2:} How to evaluate the quality of the structure of generated music pieces, especially from a \textit{harmony} perspective?

\textbf{EQ3:} Can HAT improve the quality of generated music, especially on the \textit{form} and \textit{texture}? How effective are the proposed hierarchical structure-enhanced mechanisms?

%
%

\subsection{Dataset}\label{sec:dataset}
We adopt POP909~\cite{pop909-2020}\footnote{\url{https://github.com/music-x-lab/POP909-Dataset}} as our experimental dataset because it contains sufficient annotations information of \textit{structure}. There are 909 MIDI files of pop songs of the dataset, in which every song is arranged by professional musicians as piano. For \textit{texture}, there are three tracks in every MIDI, which can respectively represent the typical pop music texture elements, Primary Melody (PM), Secondary Melody (SM), and Harmonic and Rhythmic Support (HRS)~\cite{wiki-texture}, which is shown in Figure~\ref{fig:intro-case-score}. For \textit{form}, other researches have annotated the phrase-level structure (\textit{melodic} or \textit{non-melodic} phrases and their bar-level durations) on POP909~\cite{dai-structure-analysis-2020}\footnote{\url{https://github.com/Dsqvival/hierarchical-structure-analysis}}. 

When preprocessing, we select the songs that are 4/4 time signature as our training data, including 857 MIDI files. 
We quantize the duration and the beat positions under the resolution of the $16^{th}$ note. And we stay only the first tempo value as the tempo of every song, to reduce the burden of model learning.

\begin{table*}[t]
\centering
\caption{Results of the Next Token Prediction task. (The best and the second best results of every column are \textbf{bold} and \textit{italic}.)}\label{tab:results-understanding}
\resizebox{1.5\columnwidth}{!}{%
\begin{tabular}{lcccccccc}
\toprule
\multicolumn{1}{c}{\multirow{2}{*}{\textbf{Model}}} & \multicolumn{4}{c}{\textbf{Accuracy}} & \multicolumn{4}{c}{\textbf{Mean Square Error} ($\downarrow$)} \\ \cmidrule(lr){2-5} \cmidrule(lr){6-9}
\multicolumn{1}{c}{} & \textbf{Note} & \textbf{Chord} & \textbf{Phrase} & \textbf{Avg.} & \textbf{Note} & \textbf{Chord} & \textbf{Phrase} & \textbf{Avg.} \\ \midrule
CP-Transformer~\cite{CP-Transformer-2021} & 0.406 & 0.368 & - & 0.387 & 0.132 & 0.135 & - & 0.134 \\
Music Transformer~\cite{MusicTransformer-2019} & \textit{0.587} & 0.488 & 0.256 & 0.444 & \textit{0.078} & 0.084 & 0.121 & 0.094 \\ \midrule
HAT-base & 0.485 & 0.417 & 0.228 & 0.377 & 0.099 & 0.099 & 0.124 & 0.107 \\
HAT-base w/ Form & 0.564 & 0.500 & \textit{0.309} & \textit{0.458} & 0.082 & \textit{0.082} & \textbf{0.116} & \textit{0.093} \\
HAT-base w/ Texture & 0.571 & \textit{0.503} & 0.268 & 0.447 & 0.081 & 0.084 & 0.122 & 0.096 \\
\textbf{HAT} & \textbf{0.594} & \textbf{0.518} & \textbf{0.323} & \textbf{0.478} & \textbf{0.076} & \textbf{0.080} & \textbf{0.116} & \textbf{0.090} \\ \bottomrule
\end{tabular}%
}
\end{table*}

\begin{figure*}[t]
     \centering
     \begin{subfigure}[b]{0.343\textwidth}
         \centering
         \includegraphics[width=\textwidth]{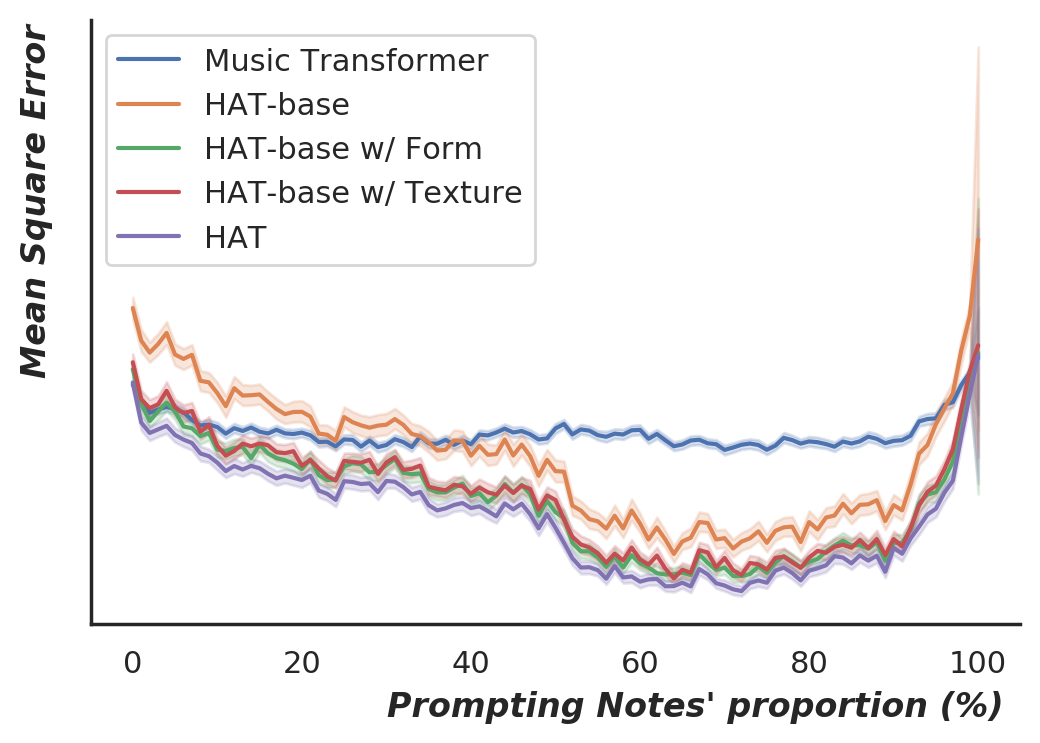}
         \caption{Note}
         \label{fig:prediction-note}
     \end{subfigure}
     \hfill
     \begin{subfigure}[b]{0.323\textwidth}
         \centering
         \includegraphics[width=\textwidth]{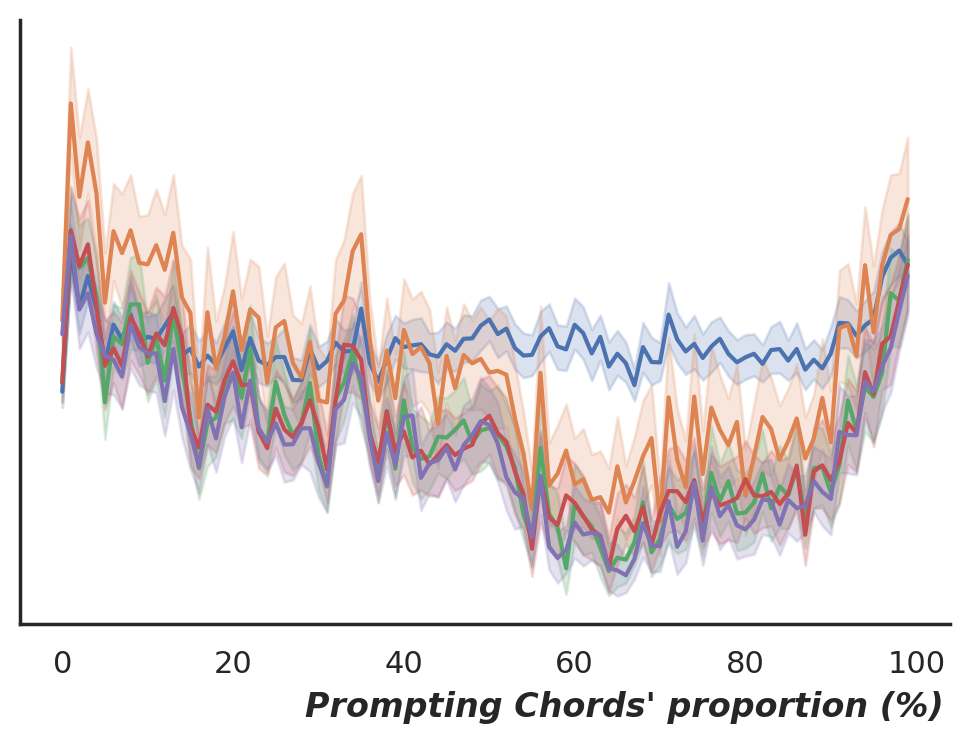}
         \caption{Chord}
         \label{fig:prediction-chord}
     \end{subfigure}
     \hfill
     \begin{subfigure}[b]{0.323\textwidth}
         \centering
         \includegraphics[width=\textwidth]{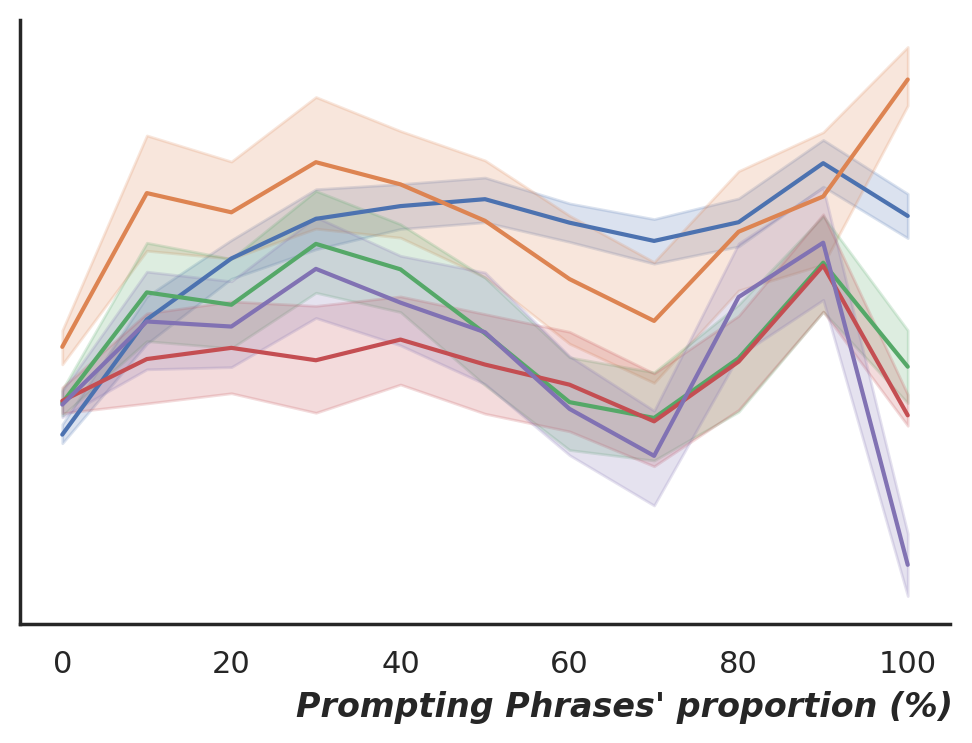}
         \caption{Phrase}
         \label{fig:prediction-phrase}
     \end{subfigure}
     \caption{The trends of the next token prediction's MSE as the prompts' lengths increase (i.e. as the music "listened" by models is going). The horizontal axes of Figure~\ref{fig:prediction-note}, ~\ref{fig:prediction-chord}, and ~\ref{fig:prediction-phrase} indicate the proportions of Note, Chord, and Phrase tokens in the prompts.}
     \label{fig:token-prediction}
\end{figure*}

\subsection{Experimental Setup}

\subsubsection{Compared Methods}\label{sec:compared-methods}

We consider the two end-to-end methods that can handle the full-song-length pop music as baselines: 
\begin{itemize}
	\item \textbf{Music Transformer}~\cite{MusicTransformer-2019} (\textbf{HAT-base w/ relative attention}): The authors employ the improved \textbf{relative attention} to model the relationship between the \textit{notes} to exhibit long-term structure. It is explained that the original tokenize methods of Music Transformer is hard to handle the full-song-length pop music~\cite{PopMusicTransformer-REMI-2020,CP-Transformer-2021}. So we adopt the \textbf{relative attention} in our proposed basic Transformer of HAT (\textbf{HAT-base}, which will be described later on).
	\item \textbf{CP-Transformer}~\cite{CP-Transformer-2021}: It is one of the SOTA end-to-end methods for pop music generation. The authors propose the "Compound Word" tokenization to compress the input sequence length prominently for Transformer-based models.
\end{itemize}
 \vspace{-2pt}

Moreover, we also adopt the three variants of HAT to study the effectiveness of our proposed components.

\begin{itemize}
	\item \textbf{HAT-base (HAT w/o Structure-enhanced)}: To verify the necessity of the HSE Module, we remove the whole module and get the variant HAT-base, which can be treated as a basic Transformer using HAT's music tokenization.
	\item \textbf{HAT-base w/ Form}: In the HSE module, we only group all the \textit{phrase}-level tokens and input them into the Form Transformer. In other words, we only enhance the \textit{form} structure on the basis of HAT-base.
	\item \textbf{HAT-base w/ Texture}: In the HSE module, we only group all the \textit{chord}-level tokens and input them into just one Texture Transformer. It means that we only enhance the \textit{texture} structure on the basis of HAT-base.
\end{itemize}

\vspace{-2pt}
\subsubsection{Implementation Details}\label{sec:implementation}
In HAT, the numbers of layers and heads are 6 and 8 for Song Transformer, 6 and 4 for Texture Transformer, and 12 and 8 for Form Transformer. The max sequence length is 2560 (Song Transformer), 60 (Texture Transformer), and 30 (Form Transformer). The embedding dim $D_S$ is 512. When training, the loss weights $\lambda_c$ are 5 (for Type and Bar), 10 (for Tempo and Phrase), and 1 (for the others). The batch size is 8, the learning rate is $10^{-4}$, and we use Adam~\cite{Adam-2015} for optimization with $\epsilon = 10^{-8}$, $\beta_1 = 0.9$, $\beta_2 = 0.999$. For generation, we adopt the models of training loss at 0.05 level to generated pieces.

\vspace{-5pt}
\subsection{Performance of Music Understanding (EQ1)}\label{sec:expt-understanding}

To evaluate HAT's ability to understand the constitution of the music, especially the musical structure, we employ the \textbf{Next Token Prediction} -- a music understanding task. Specifically, given the prompt $\mathbf{P} = [t_1, t_2, .., t_{L_\mathbf{P}}]$ (where the $L_\mathbf{P}$ is the length of the prompt), the model is expected to predict the next musical token $t_{L_\mathbf{P} + 1}$. And we adopt Accuracy and Mean Square Error (MSE) as the criteria. 

In Table~\ref{tab:results-understanding}, we report the evaluation results of three particular types of tokens: Note, Chord, and Phrase. It can be observed that:
\begin{itemize}
    \item  The effectiveness of our music tokenization for phrases: compared to CP-Transformer~\cite{CP-Transformer-2021} that does not consider to tokenize the phrase tokens, other five models own a better understanding of music, which verifies our tokenization is proper to model the phrases' signals.
    \item The superiority of HAT: among the models, HAT has the highest accuracy and the lowest MSE on any types of tokens.
    \item The advantage of the hierarchical structure-enhanced mechanisms: on the one hand, enhancing form (HAT-base w/ Form), texture (HAT-base w/ Texture), or the both (HAT) all behave better when predicting chords and phrases than Music Transformer~\cite{MusicTransformer-2019} and HAT-base. On the other hand, form- or texture-enhanced mechanism respectively improves the understanding of phrase or chord more.
\end{itemize}

Furthermore, in order to explore the changes of the model's intellect as the music is going, we visualize the trends of its prediction's MSE as the prompt's length increases for the five models using our tokenization (Figure~\ref{fig:token-prediction}). We can see that for Note (Figure~\ref{fig:prediction-note}) and Chord (Figure~\ref{fig:prediction-chord}), the MSEs of HAT and its three variants descend first, 
but then ascend until the music ends. This is mostly likely because that the intellects of these models increase as they gradually perceive the first-half music usually repeat. On the contrary, modeling the relative attention between notes (Music Transformer~\cite{MusicTransformer-2019}) can not understand more as the music goes. However, when the music is coming to the end, all the models are helpless since the music at that time often doesn't appear before.

For Phrase (Figure~\ref{fig:prediction-phrase}): (1) Initially, the five models understand little, which indicates that they don't know where the music will go by listening just the beginning of the music - maybe a \textit{intro} or a \textit{verse}. (2) Gradually, they behave better and better during the first-half music, which seems that they have perceived the repeating \textit{phrases} and the similar \textit{sections}. (3) Yet after that time, the models' intellects tend to be worse again. We speculate that there appears to be some new sections like \textit{post-chorus}~\cite{wiki-postchorus}. (4) When the music has progresses about 90\%, however, except for HAT-base, the other four structure-enhanced models (especially HAT) are conscious of the music is approaching to the end. Combining the analysis of Note and Chord before, in a word, these four models know more "when to end", but less "how to end".

\subsection{Evaluation Metrics of Music Generation (EQ2)}

\subsubsection{Objective Evaluation}\label{sec:obj-eval}


For algorithmic composition, it is still very hard to measure the quality of the generated pieces' \textit{structure}, especially by the quantitative analysis.
In the paper, we explore to develop two metrics to evaluate the \textit{texture} and \textit{form} structure of music. Particularly, we research to assess the two from the perspective of \textit{harmony}.

\paragraph{\textbf{Accompaniment Groove Stability}}

For \textit{texture}, we design the Accompaniment Groove Stability (AGS) to measure the stability of the grooves between the accompaniment textures. As the Figure~\ref{fig:intro-case-score} shows, the accompaniment textures of pop music usually appear in chords, and the grooves of the adjacent chords of the same durations are highly similar. On this observation, we formulate the grooves of the $i^{th}$ chord as $\mathbf{g}_i$, which is a binary vector $\in \mathds{R}^{D_{u}}$, where $D_{u}$ is the dimension of the chord's duration
. Then we can calculate the \textbf{AGS} as follows:

\begin{equation}
\begin{split}
	\textbf{AGS}_{i, i+1} &= 1 - \frac{{\rm \textbf{Sum}}( {\rm \textbf{XOR}}(\mathbf{g}_i, \mathbf{g}_{i+1}))}{{\rm \textbf{Sum}}({\rm \textbf{OR}}(\mathbf{g}_i, \mathbf{g}_{i+1}))},  \\
	\textbf{AGS} &= {\rm \textbf{Avg}}~(\sum_i \textbf{AGS}_{i, i+1}),
\end{split}
\label{eq:ags}
\end{equation}
where $\textbf{AGS}_{i, i+1}$ means the $i^{th}$ chord and its next adjacent $(i+1)^{th}$ chord with the same duration. ${\rm \textbf{XOR}}(\cdot,\cdot)$ and ${\rm \textbf{OR}}(\cdot,\cdot)$ are the exclusive OR and the OR operation respectively, and ${\rm \textbf{Sum}}( {\rm \textbf{OR}}(\mathbf{g}_i, \mathbf{g}_{i+1}))$ is the scaling factor. And the AGS value ranges from 0 to 1.


To obtain the grooves of the $i^{th}$ chord, $\mathbf{g}_i$, motivated by~\cite{JazzTransformer-2020}, we scan the every frame of the chord (by the resolution of $16^{th}$ note), and the value of the $j^{th}$ frame, $g_i^j = 1$ if there are notes onsetting in the frame, otherwise $g_i^j = 0$. Besides, we remove those adjacent chord pairs that $\mathbf{g}_i = \mathbf{0}$ and $\mathbf{g}_{i+1} = \mathbf{0}$ during calculating AGS.

Ideally, the stabler of the accompaniment grooves (during the adjacent chord pairs) among the entire piece, the higher AGS is, which means the \textit{texture} of the piece is better structured.

\paragraph{\textbf{Chord Progression Realism}}

For \textit{form}, we design the Chord Progression Realism (CPR) to evaluate the realism of the \textit{chord progressions} of a generated piece. As the Figure~\ref{fig:intro-case-cp} shows, the chord progression of a real piece often repeats within the phrases and the sections, resulting in a low Chord Progression Irregularity (CPI) that proposed in~\cite{JazzTransformer-2020}. 
However, we found only CPI~\cite{JazzTransformer-2020} not suitable enough to evaluate the quality (or the realism) of the chord progressions of the music. For example, if there is only one chord repeating during the entire piece, the CPI will be very low, but the music maybe like several notes' meaningless loop. 
 
 For pop music, within the \textbf{repeating} outline of chord progressions, there are also rich appropriate \textbf{variations} featuring the musical details and dynamics. Based on that, therefore, we propose the Chord Progression Variation Rationality (CPVR) to measure the rationality when the chords change. Furthermore, we balance CPI (i.e., \textit{stability}) and CPVR (i.e., \textit{diversity}) to obtain the final CPR value of a piece:
 
 \begin{equation}
	\begin{split}
		\textbf{CPR} &= \lambda \textbf{CPI} + (1 - \lambda) \textbf{CPVR}.
	\end{split}
\label{eq:CPR}
\end{equation}
where $\lambda$ is the hyperparameter of trade-off weight. In the experiments, we simply set $\lambda = 0.5$.

 To calculating CPVR, when a new \textit{unique n-grams chords} appears, we measure the probability of the appearance of this variation under the current harmony context. Given the $n$-grams chord progression is $C_i^{i+n}$, where $i$ is the index of the first chord, we traverse all the $C_i^{i+n}$ from $i=1$, and when a new unique $n$-grams appear, we define $V_i$ as the variation rationality value:

\begin{equation}
	V_i = \textbf{p}~(C_{i}^{i+n} | C_{i}^{i-1+n}),
\end{equation}
where $\textbf{p}(\cdot|\cdot)$ is the conditional probability, and we use the appearance frequency during the real data (i.e., all the pieces of POP909~\cite{pop909-2020}) to get the approximation. Finally, the CPVR value of a piece is obtained by averaging all the $V_i$: $\textbf{CPVR} = \textbf{Avg}(\sum_i V_i)$.


It is noted that the ranges of CPI, CPVR, and CPR are all from 0 to 1. Ideally, a piece with a good \textit{form} will own a high CPR value.

\subsubsection{Subjective Evaluation}

Qualitative evaluation by humans is always an important evaluation metric and is adopted widely~\cite{XiaoIceBand-2018,MusicTransformer-2019,PopMNet-2020,JazzTransformer-2020,CP-Transformer-2021} in music generation. In the paper, we invited 15 volunteers to conduct the following human study. The volunteers are from the music conservatory or in the music production industry. All of them are good at at least one instrument and are engaged in the musical activity for over 5 hours every day. Firstly, the subjects need to score the \textbf{Overall Performance (OP)} of the music:

\begin{itemize}
	\item \textbf{Melody (M)}: Does the melody sound beautiful?
	\item \textbf{Groove (G)}: Does the music sound fluent and pause suitably? Is the groove of the music unified and stable?
\end{itemize}

Next, the subjects are asked to evaluate the quality of \textit{texture} and  \textit{form} respectively. For \textit{texture}, there are two metrics:

\begin{itemize}
	\item \textbf{Primary Melody (PM)}: Is there a distinct and clear primary melody line in the music? Is the primary melody easy to remember? Is it suitable for singing with lyrics?
	\item \textbf{Consonance (CO)}: Do the several layers of sound balance and combine organically? Is the composition of individual sounds harmonious?
\end{itemize}

For \textit{form}, there are also two metrics:

\begin{itemize}
	\item \textbf{Coherence (C)}: Are the transitions and the developments between the contiguous \textit{phrases} natural and coherent?
	\item \textbf{Integrity (I)}: Does the music own the complete \textit{section} structure, such as intro, verse, chorus, bridge, and outro? Are the boundaries between the sections clear?
\end{itemize}

\subsection{Performance of Music Generation (EQ3)}\label{sec:eval-generation}

\subsubsection{Results of Objective Evaluation}

\begin{table}[t]
\centering
\caption{Results of the objective evaluation. The best results of every column (except those from Real) are \textbf{bold}, and the second best results are \textit{italic}.}
\label{tab:eval-obj}
\resizebox{\columnwidth}{!}{%
\begin{tabular}{lcccc}
\toprule
\multicolumn{1}{c}{\multirow{3}[3]{*}{\textbf{Model}}} & \textbf{Texture} & \multicolumn{3}{c}{\textbf{Form}} \\ \cmidrule(lr){2-2} \cmidrule(lr){3-5}
\multicolumn{1}{c}{} & \multirow{2}{*}{\textbf{AGS}} & \multicolumn{3}{c}{\textbf{CPR}} \\ \cmidrule(lr){3-5}
\multicolumn{1}{c}{} & & \textbf{2-grams} & \textbf{3-grams} & \textbf{4-grams} \\ \midrule
Real & 0.572 & 0.504 & 0.564 & 0.551 \\ \midrule
CP-Transformer~\cite{CP-Transformer-2021} & 0.193 & 0.312 & 0.250 & 0.132 \\
Music Transformer~\cite{MusicTransformer-2019} & 0.256 & 0.413 & 0.384 & 0.267 \\ \midrule
\textbf{HAT-base} & 0.382 & 0.403 & 0.369 & 0.264 \\
\textbf{HAT-base w/ Form} & 0.422 & \textit{0.439} & \textit{0.421} & 0.307 \\
\textbf{HAT-base w/ Texture} & \textit{0.456} & 0.434 & 0.417 & \textit{0.310} \\
\textbf{HAT} & \textbf{0.474} & \textbf{0.447} & \textbf{0.435} & \textbf{0.320} \\
\bottomrule
\end{tabular}%
}
\end{table}

To evaluate the generated pieces on the proposed objective metrics, we produced 100 pieces from scratch for each model. Also, we adopt all the human 857 pieces (our training data) as a "Real" model for a comparison. We obtain every model's score by averaging its pieces' scores. 

The results of the objective evaluation are exhibited in Table~\ref{tab:eval-obj}. It is observed that: (1) The Real model owns the highest AGS and CPRs, though whose scores are only 0.5-0.6. It indicates that our proposed AGS and CPR can be considered as reasonable \textit{descriptive} metrics at least, which can measure the generator's ability in style imitation; (2) For \textit{texture}, HAT is much significantly better than the two baselines on AGS (Music Transformer: $p = 0.0022$; CP-Transformer: $p = 6.59 \times 10^{-5}$; one-tailed $t$-test
); (3) For \textit{form}, on the one hand, HAT is significantly better on any grams of CPR. On the other hand, when the gram increases, the CPR values of any algorithmic music all drop a lot, but that is barely seen for real pieces. It reveals the challenge of generating longer and more abstract structures. (4) Among the HAT's three variants, it can be seen that based on HAT-base, the form or texture enhancement respectively behaves quite well on AGS or CPR. That verifies the effectiveness of our proposed hierarchical structure-enhanced strategies. 

\subsubsection{Results of Subjective Evaluation}

\begin{table}[t]
\centering
\caption{Results of the subjective evaluation. \textbf{OP}: Overall Performance, \textbf{M}: Melody, \textbf{G}: Groove, \textbf{PM}: Primary Melody, \textbf{CO}: Consonance, \textbf{C}: Coherence, \textbf{I}: Integrity.}
\label{tab:eval-sub}
\resizebox{\columnwidth}{!}{%
\begin{tabular}{lccccccc}
\toprule
\multicolumn{1}{c}{\multirow{2}[3]{*}{\textbf{Model}}} & \multicolumn{2}{c}{\textbf{OP}} & \multicolumn{2}{c}{\textbf{Texture}} & \multicolumn{2}{c}{\textbf{Form}} & \multirow{2}{*}{\textbf{Avg.}} \\ \cmidrule(lr){2-3} \cmidrule(lr){4-5} \cmidrule(lr){6-7}
\multicolumn{1}{c}{} & \multirow{1}{*}{\textbf{M}} & \multirow{1}{*}{\textbf{G}} & \multirow{1}{*}{\textbf{PM}} & \multirow{1}{*}{\textbf{CO}} & \multirow{1}{*}{\textbf{C}} & \multirow{1}{*}{\textbf{I}} & \\ \midrule
CP-Transformer~\cite{CP-Transformer-2021} & 0.356 & 0.356 & 0.385 & 0.403 & 0.419 & 0.380 & 0.383 \\
Music Transformer~\cite{MusicTransformer-2019} & 0.417 & 0.375 & \textbf{0.700} & 0.562 & 0.550 & 0.375 & 0.496 \\ \midrule
HAT-base & 0.267 & \textit{0.550} & \textit{0.680} & 0.400 & 0.400 & 0.450 & 0.458 \\
HAT-base w/ Form & \textbf{0.638} & 0.504 & 0.511 & \textit{0.641} & \textit{0.557} & \textit{0.574} & \textit{0.571} \\
HAT-base w/ Texture & 0.436 & 0.477 & 0.514 & 0.539 & 0.538 & 0.504 & 0.501 \\
\textbf{HAT} & \textit{0.592} & \textbf{0.552} & 0.598 & \textbf{0.661} & \textbf{0.585} & \textbf{0.618} & \textbf{0.601} \\
\bottomrule
\end{tabular}%
}
\end{table}

For subjective evaluation, we generated 10 pieces from scratch for each model and ask our volunteers to give a mark for all the pieces. The source of pieces are concealed from the subjects and we guarantee that each piece is rated by 3 different subjects. The results of the subjective evaluation are displayed in Table~\ref{tab:eval-sub}, where we scaled the scores and every value in it lies from 0 to 1. We can see that the general quality (the average score 0.601) and the \textit{form} quality of HAT's pieces are the best, and those of HAT-base w/ Form are the second best.





\section{Case Study}

\begin{figure}[t]
     \centering
     \begin{subfigure}[b]{0.233\textwidth}
         \centering
         \includegraphics[width=\textwidth]{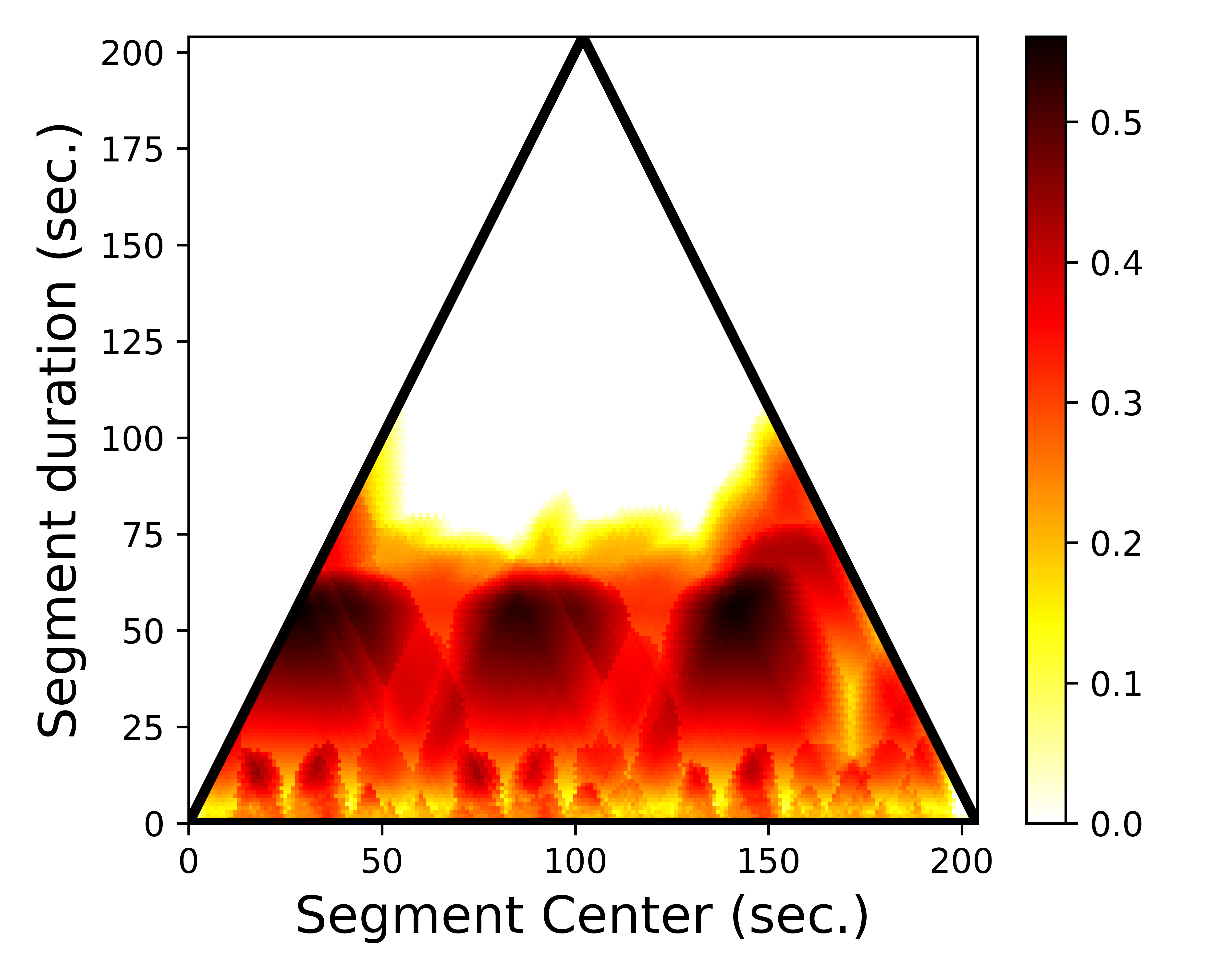}
         \caption{Real piece}
         \label{fig:case-real}
     \end{subfigure}
     \hfill
     \begin{subfigure}[b]{0.233\textwidth}
         \centering
         \includegraphics[width=\textwidth]{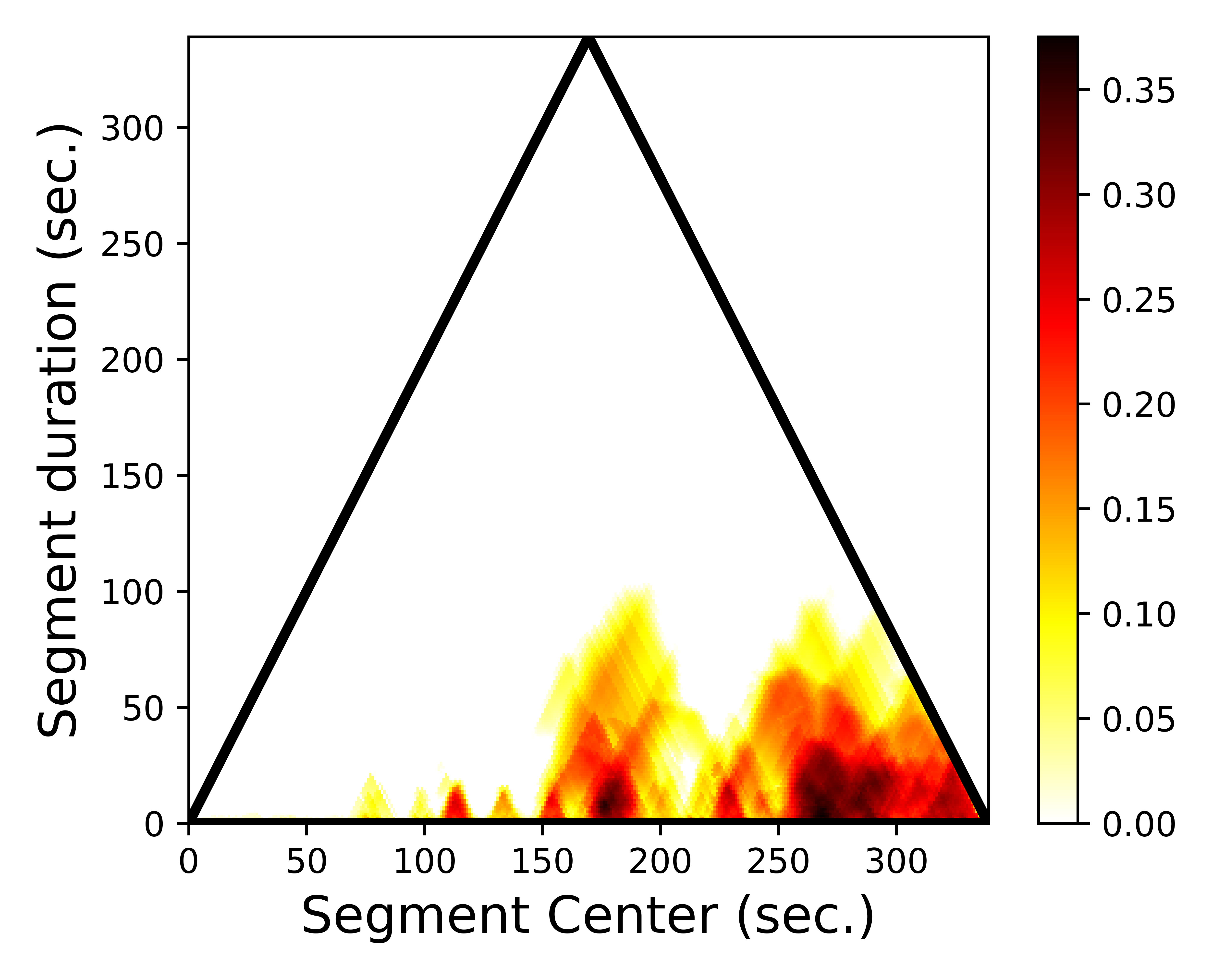}
         \caption{Music Transformer}
         \label{fig:case-music-transformer}
     \end{subfigure}
     \hfill
     \begin{subfigure}[b]{0.233\textwidth}
         \centering
         \includegraphics[width=\textwidth]{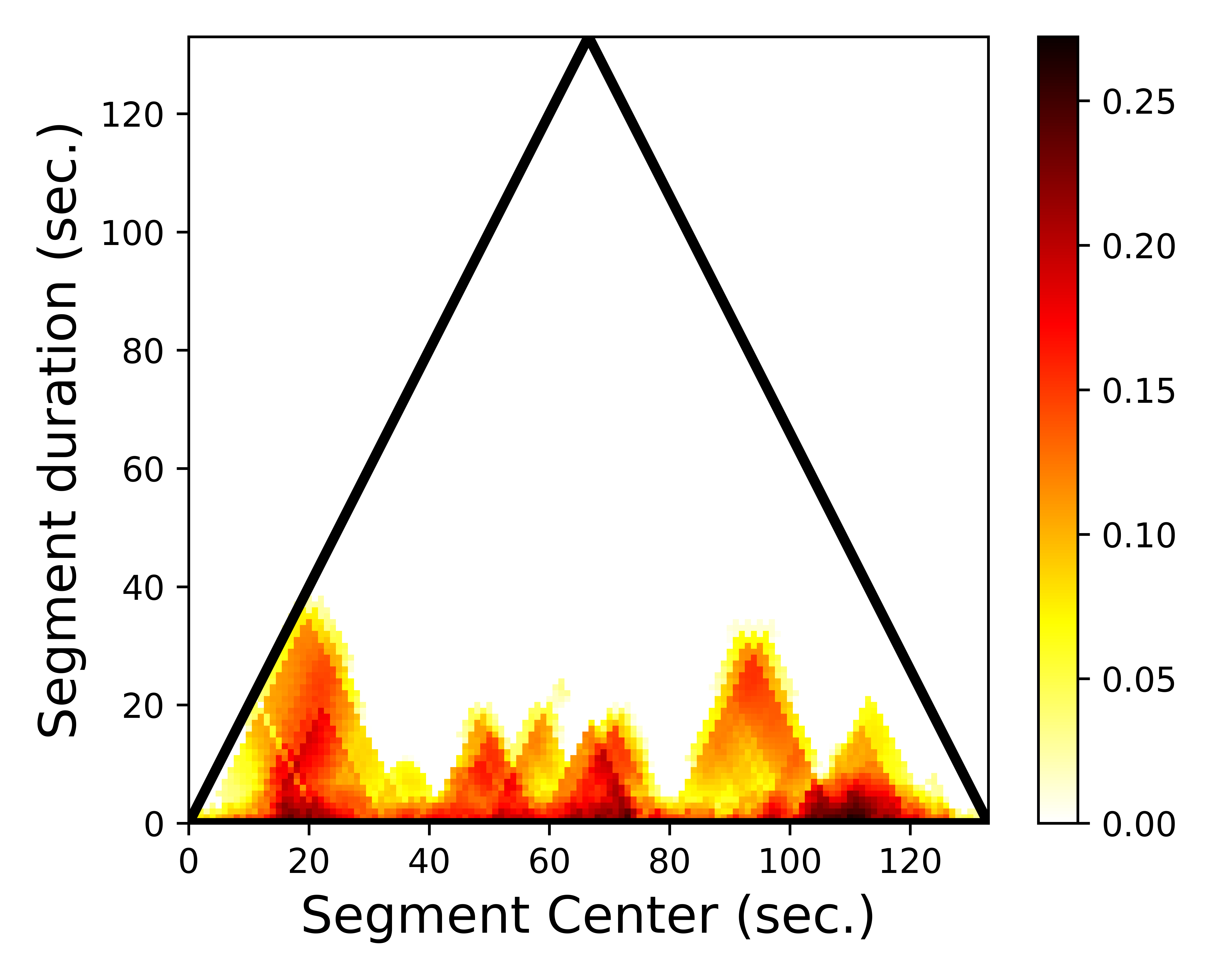}
         \caption{CP-Transformer}
         \label{fig:case-cp-transformer}
     \end{subfigure}
     \hfill
     \begin{subfigure}[b]{0.233\textwidth}
         \centering
         \includegraphics[width=\textwidth]{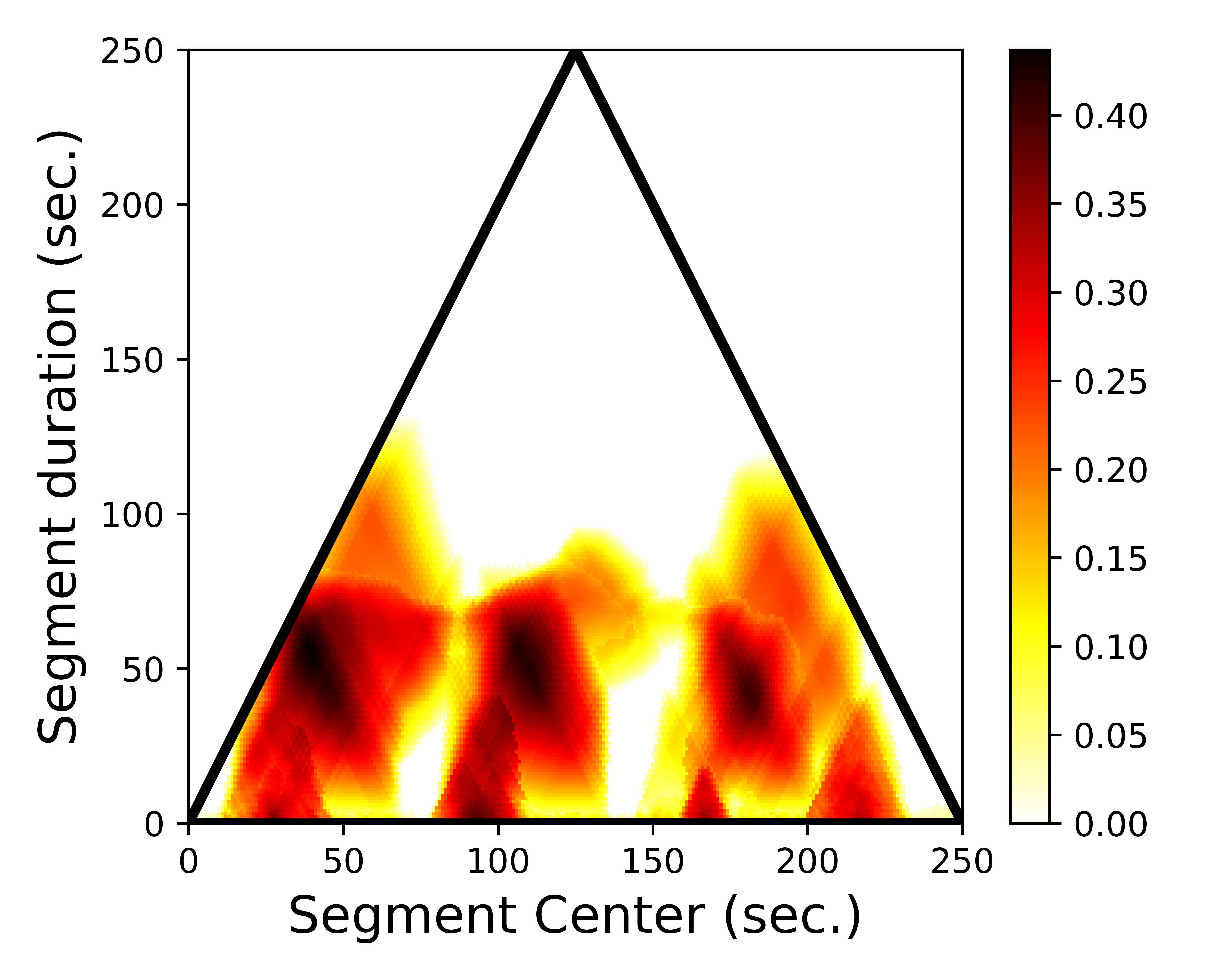}
         \caption{\textbf{HAT}}
         \label{fig:ca dse-HAT}
     \end{subfigure}
     \caption{The fitness scape plots of the real piece (the Chinese pop song, \textit{Guang Yin De Gu Shi}), and other three generated pieces prompted by the \textit{intro} of the real piece.}
     \label{fig:case}
\end{figure}

To figure out the characteristics of the pieces generated by HAT, we visualize the \textit{fitness scape} of the structure segments that are applied in the fields of Music Structure Analysis~\cite{fitness-plot-2012,FMP-book-2015}. In Figure~\ref{fig:case}, we can see that: (1) compared with the two baseline-generated pieces, the HAT-generated piece owns the less but the longer duration segments, which is similar to the real piece; (2) compared with the real piece, the HAT's piece already has the three main segments like the real one, but it lacks the musical details between the structure segments (appearing as blank in the small segments). It reveals that although HAT has been capable of imitating the outline structure of the real music, it is still too hard for it to polish and refine the generated pieces to pursue a real work of art.

\section{Conclusion and Future Work}

In this paper, we propose the \textit{harmony}-aware learning for structure-enhanced pop music generation. Bridged by the harmony, we combine the \textit{texture} and the \textit{form} structure organically, and we design the HAT for their joint learning. Both experimental results of music understanding and generation verify the significant effectiveness of the HAT.
In the future work, on the one hand, we will explore new methods to polish and refine the musical details of generated pieces. On the other hand, based on the generated symbolic music, we will research on the \textit{performance} generation, hoping to merge the human performance techniques into the generated music.

\section*{Acknowledgement}
We thank Ziniu Li, Ke Xue, Qiang Sheng, and the anonymous reviewers for their insightful comments and suggestions. We appreciate the efforts of all the volunteers during the subjective evaluation.

\bibliographystyle{ACM-Reference-Format}
\bibliography{sample-sigconf}

\clearpage

\appendix


\section{Generating a piece from the scratch}

In Section 3.6, we describe the generating procedure of HAT from the scratch. The detailed algorithm can be seen in Algorithm~\ref{alg:generate}. 

\section{Dataset of Music Understanding}
For the next token prediction task (Section 4.3), we split the 857 MIDIs (Section 4.1) into training (773 MIDIs), validation (41 MIDIs), and testing sets (43 MIDIs). We select the best model on the validation set, and exhibit the results of the testing set (Table 2 and Figure 3).

\section{Supplementary Implementation Details}

\subsection{Computing Platform}
All experiments in the paper are conducted on NVIDIA V100 GPUs with PyTorch~\cite{pytorch-2019}. And we employ the library developed by~\cite{fast-transformer-2020}\footnote{\url{https://github.com/idiap/fast-transformers}. This library is originally developed for ``fast attention"~\cite{fast-transformer-2020} for transformers. But we only use its implementation of the basic Transformer~\cite{Transformer-2017} during all the experiments.} for the implementation of transformer.

\subsection{HAT} In Section 3.6.2, we use the sampling function $\textbf{Sampling}_c(\cdot)$ for the different category $c$ of the tokens during generating. For the hyperparameters, we follow the setting of~\cite{CP-Transformer-2021} for all the categories except Phrase (that is not modeled in~\cite{CP-Transformer-2021}). And we employ the sampling function with $\tau = 1.0$ and $\rho = 0.99$ for Phrase.

\subsection{HAT's three variants} All the hyperparameters of them are the same to HAT. And we use the models of training loss at 0.05 level for generating, following HAT.

\subsection{Music Transformer} As is mentioned in Section 4.2.1, we adopt \textbf{HAT-base w/ relative attention} for the implementation of Music Transformer. We reproduce the \textbf{relative attention} based on some public codes\footnote{\url{https://github.com/jason9693/MusicTransformer-pytorch}}. And we use the models of training loss at 0.05 level for generating, following HAT.

\subsection{CP-Transformer} We follow the official implementation of CP-Transformer~\cite{CP-Transformer-2021}\footnote{\url{https://github.com/YatingMusic/compound-word-transformer}}. We merge the three-track training data into single-track data, because the original tokenize method of ``Compound Word" is proposed on the single-track setting. To be fair with HAT, we use the basic Transformer~\cite{Transformer-2017} but rather than linear Transformer~\cite{fast-transformer-2020} for the backbone architecture.

\subsection{Objective Evaluation}

\subsubsection{Chord Detection}
In Section 4.4.1, the proposed AGS and CPR are both dependent on the chord signals. In order to get the chord annotations, we use the chord detection tool\footnote{\url{https://github.com/joshuachang2311/chorder}} of~\cite{CP-Transformer-2021} to extract the chords of the evaluated pieces.

\subsubsection{Chord Progression Irregularity (CPI)}

To calculate CPI~\cite{JazzTransformer-2020} (Equation 12), the $n$-grams chord progression is notated as $C_i^{i+n}$, where $i$ is the index of the first chord, then we can obtain CPI value as follows:

\begin{equation*}
	\textbf{CPI} = \frac{{\rm \textbf{Count}}(\{C_i^{i+n}~|~i = 1, 2, ..., L_c - n+1\})}{L_c - n + 1},
\end{equation*}
where $\textbf{Count}(\cdot)$ means the size of the set, and $L_c$ is the number of the chords of the music.

\renewcommand{\algorithmicrequire}{\textbf{Input:}}
\renewcommand{\algorithmicensure}{\textbf{Output:}}
\algdef{SE}[DOWHILE]{Do}{doWhile}{\algorithmicdo}[1]{\algorithmicwhile\ #1}%

\begin{algorithm}[t]
\caption{Generate a music piece from scratch.}\label{alg:generate}
\begin{algorithmic}[1]
\Require $\mathbf{x}_1 = {\rm \texttt{<BOS>}}$
\Ensure $\hat{\mathbf{y}} = [\hat{\mathbf{y}}_1, \hat{\mathbf{y}}_2, ..., \hat{\mathbf{y}}_{L_G}]$

\State $i \gets 0$ \Comment{the token index}
\State $p_i \gets 0$ \Comment{the phrase index}
\State $p_i^j \gets 0$ \Comment{the chord index of the phrase}
\\
\State $\mathbf{T}_{p_i^j - 1} \gets \mathbf{0}$ \Comment{the output of Texture Transformer}
\State $\mathbf{TF}_{p_i - 1} \gets \mathbf{0}$ \Comment{the output of Form Transformer}
\\
\Do
\State $i \gets i + 1$
\State Calculate $\mathbf{E}_{\mathbf{x}_i}$ and $\mathbf{S}_{\mathbf{x}_i}$ using Equation 1-2.
\If{$\mathbf{x}_i^{tp} = phrase$}
	\State $\mathbf{S}_{p_i} \gets \mathbf{S}_{\mathbf{x}_i}$
	\State $\mathbf{S}_{\mathbf{x}_i} \gets \mathbf{TF}_{p_i - 1} + \mathbf{S}_{\mathbf{x}_i}$ \Comment{Refer to Equation 5}
	\State $\mathbf{S}_{p_i}^\prime \gets \mathbf{S}_{\mathbf{x}_i}$
	\If{$p_i \neq 0$}
	\State Given $\mathbf{T}_{p_i^j - 1}$, calculate $\mathbf{TF}_{p_i - 1}$ using Equation 4.
	\EndIf
	\State $p_i \gets p_i + 1$
	\State $p_i^j \gets 0$
	\State $\mathbf{T}_{p_i^j - 1} \gets \mathbf{0}$
\ElsIf{$\mathbf{x}_i^{tp} = chord$}
	\State $\mathbf{S}_{p_i^j} \gets \mathbf{S}_{\mathbf{x}_i}$
	\State $\mathbf{S}_{\mathbf{x}_i} \gets \mathbf{S}_{p_i}^\prime + \mathbf{T}_{p_i^j - 1} + \mathbf{S}_{\mathbf{x}_i}$ \Comment{Refer to Equation 6}
	\State Given $\mathbf{S}_{p_i^j}$ and $\mathbf{S}_{p_i}$, calculate $\mathbf{T}_{p_i^j - 1}$ using Equation 3.
	\State $p_i^j \gets p_i^j + 1$
\EndIf
\State $\mathbf{S}_{\mathbf{x}_i}^\prime \gets \mathbf{S}_{\mathbf{x}_i}$
\State Given $\mathbf{S}_{\mathbf{x}_i}^\prime$, calculate $\mathbf{S}_{\mathbf{x}_i}^{\prime\prime}$ using Equation 7.
\State Given $\mathbf{S}_{\mathbf{x}_i}^{\prime\prime}$, calculate $\hat{\mathbf{y}}_i$ using Equation 8, where we only replace $\mathbf{y}_i^{tp}$ with $\hat{\mathbf{y}}_i^{tp}$.
\doWhile $\hat{\mathbf{y}}_i \neq  {\rm \texttt{<EOS>}}$
\end{algorithmic}
\end{algorithm}

\end{document}